\documentclass[aps,pra,twocolumn,reprint,showpacs,groupedaddress]{revtex4-1}
\usepackage{amssymb,amsmath,amsfonts} \usepackage{epsfig,graphicx}

\usepackage{latexsym}
\usepackage{stmaryrd}
\usepackage[table]{xcolor}
\usepackage[colorlinks,citecolor=blue]{hyperref}
\usepackage{bm}
\usepackage{bbm}
\usepackage{color}
\usepackage[english]{babel}
\usepackage{subfigure}
\usepackage{natbib}
\usepackage{appendix}
\newcommand{\ool}{\Omega/\omega_0 \rightarrow \infty} 
\newcommand{\oo}{\Omega/\omega_0}
\newcommand{\adag}{a^{\dagger}}
\newcommand\ket[1]{\left|\textstyle{#1}\right\rangle}
\newcommand\bra[1]{\left\langle\textstyle{#1}\right|}
\newcommand\braket[1]{\left\langle\textstyle{#1}\right\rangle}

\begin{document}
\title{Excited-state quantum phase transition in the Rabi model}
\author{Ricardo Puebla, Myung-Joong Hwang, and Martin B. Plenio}
\affiliation{Institut f\"{u}r Theoretische Physik and IQST,
  Albert-Einstein Allee 11, Universit\"{a}t Ulm, 89069 Ulm, Germany}

\begin{abstract}
The Rabi model, a two-level atom coupled to a harmonic oscillator, can undergo a second-order quantum phase transition (QPT) [M. -J. Hwang {\it et al}, Phys. Rev. Lett. 115, 180404 (2015)]. Here we show that the Rabi QPT accompanies critical behavior in the higher energy excited states, i.e., the excited-state QPT (ESQPT). We derive analytic expressions for the semiclassical density of states, which shows a logarithmic divergence at a critical energy eigenvalue in the broken symmetry (superradiant) phase. Moreover, we find that the logarithmic singularities in the density of states leads to singularities in the relevant observables in the system such as photon number and atomic polarization. We corroborate our analytical semiclassical prediction of the ESQPT in the Rabi model with its numerically exact quantum mechanical solution.
 \end{abstract}


\maketitle

\section{Introduction}
\label{sec:i}

The understanding of phase transitions at zero temperature has been an intense area of research both theoretically and experimentally during the last decades~\cite{Sachdev:11,Vojta:03,Baumann:10,Baumann:11,Greiner:02,Bloch:12}. Such quantum phase transitions (QPT) describe an abrupt and non-analytic change of the ground state properties as the control parameter of a Hamiltonian is varied, and the critical value of the control parameter where the QPT occurs is called as a critical point. The critical point therefore divides a normal phase and the symmetry broken phase. While the ground state QPT concerns itself mainly with the lowest-energy sector in the energy spectrum, there is another kind of quantum criticality that appears in the higher energy sector of the spectrum, known as the excited-state quantum phase transition (ESQPT)~\cite{Heiss:02,Leyvraz:05,Cejnar:06,Cejnar:08,Caprio:08,Stransky:14}.

The ESQPT describes an abrupt change in the nature of the eigenstates and energy spectrum at a critical energy, which is generally much larger than the ground state energy, in the symmetry broken phase. Particularly, the critical energy divides the energy spectrum into two parts: (i) below the critical energy all the eigenstates have degeneracies arising from the spontaneous symmetry breaking (ii) above the critical energy, the eigenstates are non-degenerate and restore the symmetry of the Hamiltonian to that of the normal phase. Moreover, the prominent feature of the ESQPT is known to be a singularity in the density of states at the critical energy. The paradigmatic examples include the Dicke model~\cite{Dicke:54,Hepp:73,Wang:73,Emary:03,Emary:03a} and the  Lipkin-Meshkov-Glick model~\cite{Lipkin:65}, which undergo a QPT in the thermodynamic limit, but at the same time, have a finite number of collective degrees of freedom. Experimentally, these singularities in the density of states have been observed recently using microwave photonic crystals~\cite{Dietz:13,Iachello:15}. In addition, it is also worth mentioning the dynamical relevance of ESQPTs across different low-dimensional systems, which has led to predictions of rich variety of phenomena~\cite{Relano:08dec,Perez-Fernandez:11,Yuan:12,Relano:14,Bastidas:14,Engelhardt:15,Puebla:15,Kopylov:15,Lobez:16}, for example, the existence of symmetry-breaking equilibrium states~\cite{Puebla:13}.

Recently, it has been recognized that a system with a finite number of system components can undergo a second-order quantum phase transition~\cite{Hwang:15,Hwang:16}. It has been shown that the infinite dimensional Hilbert space of even a single bosonic mode, together with a strong coupling to an atom, plays a role equivalent to the thermodynamic limit achieved by an infinite number of system components, leading to the emergence of the finite-system QPT~\cite{Hwang:15,Hwang:16}. An important question in this context is then whether the finite-system ground state QPT also accompanies the ESQPT.

In the present article, we consider a simple and ubiquitous quantum system that describes the interaction between a single two-level system (TLS) and a single-mode cavity field, known as Rabi model. The Rabi model undergoes a second-order QPT in the limit of  $\ool$ and $\lambda\rightarrow\infty$ where $\Omega$ and $\omega_0$ are the characteristic frequencies of the TLS and cavity field, respectively, and $\lambda$ is the interaction strength between them~\cite{Hwang:15}. We show that the Rabi model does exhibit all the hallmarks of the ESQPT including the logarithmic divergence of the semiclassical density of states at the critical energy, the critical behavior of the mean-field observables, and the precursors of the ESQPT such as the level-clustering and the crossover from the nearly degenerate low-energy sector to the non-degenerate high-energy sector in the broken symmetry phase.

This article is organized as follows. In Sec.~\ref{sec:ii} the quantum Rabi model is introduced and the precursors of the ESQPT in the energy spectrum have been presented. After the introduction of the semiclassical limit of the Rabi model in Sec.~\ref{sec:iiia}, we present analytical and numerical analysis on the semiclassical and quantum density of states in Sec.~\ref{sec:iiib}. We show that the semiclassical and quantum density of states unveil the presence of an ESQPT, which consists in a logarithmic divergence of the density of states at a certain critical energy.  Furthermore, we show that the density of states diverges at the ground-state QPT and that its divergence is characterized by a power law. In Sec.~\ref{sec:iiic}, we show that, as a result of the ESQPT, relevant observables of the system, namely, photon number and TLS occupation, inherit the singular behavior of the density of states. In all analysis, the comparison between the semiclassical and the quantum calculations shows an excellent agreement, provided that the frequency ratio $\oo$ is large enough. Finally, we conclude our study in Sec.~\ref{sec:iv}.

\section{The Quantum Rabi Model}
\label{sec:ii}
\subsection{Hamiltonian}
The Rabi model describes the interaction of a single two-level system (TLS) with a single-mode cavity field, whose Hamiltonian reads
\begin{equation}
\label{eq:H}
H= \omega_0 a^{\dagger}a + \frac{\Omega}{2} \sigma_z - \lambda \left(
a^{\dagger} + a\right) \sigma_x.
\end{equation}
Here $a^{\dagger}$ and $a$ are the creation and annihilation operator of the cavity field, respectively, and $\sigma_{x,y,z}$ are the Pauli matrices. The cavity frequency is $\omega_0$, the transition frequency of the TLS is $\Omega$, and the coupling strength $\lambda$. The basis state is $\left|n,\sigma \right>$, a product state of a $n$-photon Fock state $\ket{n}$ and a spin state $\ket{\sigma=\uparrow(\downarrow)}$ with $\sigma_z\ket{\uparrow(\downarrow)}=\pm\ket{\uparrow(\downarrow)}$. We set $\hbar=1$ throughout the whole article. The Rabi Hamiltonian $H$ has a discrete $Z_2$ symmetry~\cite{Braak:11,Hwang:10}, that is, $[\Pi,H]=0$ where the parity operator $\Pi=e^{i\pi a^{\dagger} a}\sigma_z$. The even ($+$) and odd ($-$) parity are therefore good quantum numbers and we denote the $k$-th eigenstate in each parity subspace as $\left|\varphi_k^{\pm}\right>$, which satisfies $H\left|\varphi_k^{\pm}\right>=E_k^{\pm}\left|\varphi_k^{\pm}\right>$ and $\Pi\left| \varphi_k^{\pm}\right>=\pm \left| \varphi_k^{\pm}\right>$ where $E_k^{\pm}$ is the corresponding energy eigenvalues. We note that each parity subspace can be effectively described as a single non-linear harmonic oscillator~\cite{Hwang:10}; therefore each parity subspace consists of a single degree of freedom. 

\subsection{Second-order QPT and energy spectrum}
\begin{figure}
\includegraphics[width=0.5\linewidth,angle=-90]{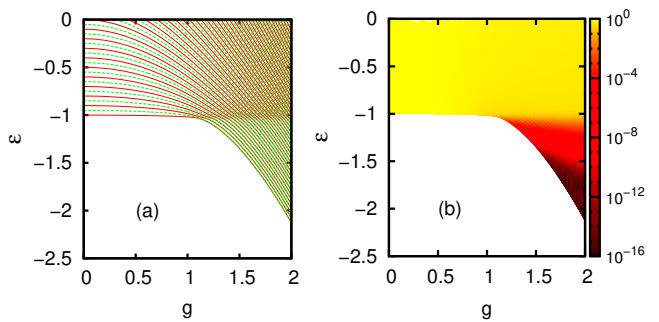}
\caption{(Color online) The quantum Rabi model (a) The energy spectrum for $\Omega/\omega_0=40$ as a function of the dimensionless coupling strength $g$. The energy eigenvalues are divided by $\Omega/2$ and denoted as $\varepsilon$. The red solid and dashed green lines correspond to the negative and positive parity eigenstates, respectively. For $g\gtrsim1$ and $\varepsilon\lesssim-1$, there is a pair of nearly degenerate ground state. (b) The phase diagram for the energy difference $\Delta_k$ as a function of the rescaled energy $\varepsilon$ and $g$. There is a sharp crossover at $\varepsilon=-1$.}
\label{fig:deg}
\end{figure}

It has recently been shown that the Rabi model undergoes a second-order QPT in the joint limit of $\ool$ and $\lambda\rightarrow \infty$ where the control parameter $g=2\lambda/\sqrt{\omega\Omega}$ is kept constant~\cite{Hwang:15}. The critical point at which the ground state energy and the order parameter become non-analytic is $g=1$. Below the critical point $g<1$ is the normal phase, where all energy eigenstates respect the $Z_2$ symmetry. Above the critical point $g>1$ is the superradiant phase, where the ground state becomes doubly degenerate, $E_0^+=E_0^-$, and they are no longer eigenstate of the parity symmetry operator, $\Pi\ket{\varphi_0^{\pm}}\neq\pm\ket{\varphi_0^{\pm}}$~\cite{Hwang:15}. The order parameter is the spontaneous coherence of the cavity field $\braket{a}=\braket{\varphi_0^\pm|a|\varphi_0^\pm}$, which is zero for the normal phase ($g<1$) and is non-zero for the superradiant phase ($g<1$).

While the second-order QPT is primarily concerned with the ground state properties, the analytical solution of the Rabi model for low-energy physics in the $\ool$ limit shows that all the low-lying eigenstates in the superradiant phase of the Rabi model are also doubly degenerate with spontaneously broken-symmetry~\cite{Hwang:15}. In other words, for $g>1$, we have $E_k^+=E_k^-$ and $\Pi\ket{\varphi_k^{\pm}}\neq\pm\ket{\varphi_k^{\pm}}$ for any finite integer $k$. This observation opens a question whether the Rabi QPT is accompanied by an ESQPT in the $\ool$ limit; that is, whether there is a critical energy $E_c$ in the energy spectrum such that for $k$ eigenstates satisfying $E_k^\pm>E_c$ the degeneracy is lifted, $E_k^+\neq E_k^-$, and the symmetry of the eigenstate is restored,  $\Pi\ket{\varphi_k^{\pm}}=\pm\ket{\varphi_k^{\pm}}$. While the critical point $g=1$ divides the normal and the superradiant (broken-symmetry) phase, the critical energy $E_c$, if it exists, divides the energy spectrum within the broken-symmetry phase into two sectors: one in which all the energy eigenstates are doubly degenerate with spontaneously broken symmetry ($E_k<E_c$) and the other in which eigenstates are non-degenerate and respect the parity symmetry as in the normal phase  ($E_k>E_c$)~\cite{Puebla:13}.

A numerically exact diagonalization of the quantum Rabi model in Eq.~(\ref{eq:H}) for a large but finite value of $\oo$ strongly suggests that there is indeed a critical energy in the superradiant phase. In Fig.~\ref{fig:deg} (a), we present the energy spectrum of the Rabi model for $\oo=40$ as a function of $g$. For convenience, we divide the energy eigenvalues $E_k$ by the absolute value of the ground state energy at $g=0$, i.e., $\varepsilon_k\equiv 2E_k/\Omega$. It is evident that there is a critical point $g\sim1$ above which the ground states as well as the low-lying eigenstates become nearly degenerate. Due to the \emph{finite-frequency} effect, which is analogous to finite-size effect in traditional QPT~\cite{Hwang:15}, the degeneracy between $\ket{\varphi_k^{+}}$ and $\ket{\varphi_k^{-}}$ for small $k$ is lifted, but the energy difference $\Delta_k=\varepsilon_k^+-\varepsilon_k^-$ is inversely proportional to $\oo$ so that it becomes very small. In Fig.~\ref{fig:deg} (b), we present a phase diagram for $\Delta_k$ as a function of the rescaled energy $\varepsilon$ and the coupling strength $g$. In the superradiant phase, $g>1$, there is a sharp crossover at $\varepsilon=-1$ from the low-energy sector ($\varepsilon<-1$) with nearly degenerate pairs of eigenstates to the high-energy sector ($\varepsilon>-1$) with well-separated energy levels. Furthermore, at the point of the crossover $\varepsilon=-1$, there occurs a level clustering [Fig.~\ref{fig:deg} (a)]; this is a precursor of the diverging semiclassical density of states at $\varepsilon=-1$ that we will show in the next section.

\section{Excited-state quantum phase transition}
\label{sec:iii}
Motivated by the observations in the previous section, here we study the semiclassical limit of the Rabi model as  the ESQPT is related to particular changes in the phase space of the semiclassical limit of the quantum system leading to singularities in the density of states~\cite{Heiss:02,Leyvraz:05,Cejnar:06,Caprio:08,Cejnar:08,Brandes:13,Ribeiro:08,Ribeiro:07,Bastarrachea:14,Stransky:14}.
\subsection{Semiclassical limit}
\label{sec:iiia}
The semiclassical limit of the Rabi model can be taken by replacing the cavity field operator  $a$ by a complex number. Although the diverging quantum fluctuation of the cavity field in the $\ool$ limit of the Rabi model shown in Ref.~\cite{Hwang:15} cannot be properly taken into account in this semiclassical approach, the mean-field values such as the ground state energy, the photon population and the atomic population of the ground state can nevertheless be described adequately by the semiclassical approach~\cite{Ashhab:10,Ashhab:13,Bakemeier:12}. This is also the case for the Dicke or Lipkin-Meshkov-Glick models in the thermodynamic limit, where ESQPT has been successfully investigated in the semiclassical limit~\cite{Leyvraz:05,Caprio:08,Brandes:13,Bastarrachea:14}.

We describe the harmonic oscillator by means of its position and momentum operators $(\hat{x},\hat{p})$, which can be written in terms of the bosonic operators as
\begin{align}
\label{eq:xp}
\hat{x} &= \frac{1}{\sqrt{2}}\left(a^{\dagger}+a \right),
\\ \hat{p} &= i\frac{1}{\sqrt{2}}\left(a^{\dagger}-a \right).
\end{align}
The semiclassical Hamiltonian can be obtained by considering the previous
operators as continuous variables $(\hat{x},\hat{p})
\rightarrow (x',p')$, that is,
\begin{align}
\label{eq:Hscl}
H_{\textrm{scl}}(x',p')/\Omega= \frac{\omega_0}{2\Omega}\left(x^{\prime 2}+p^{\prime 2}\right)+\frac{1}{2}\sigma_z -g\sqrt{\frac{\omega_0}{2\Omega}} x^{\prime}\sigma_x,
\end{align}
up to the constant energy shift of $-\frac{\omega_0}{2\Omega}$. Then, we diagonalize the spin Hamiltonian, the last two terms, of Eq.~(\ref{eq:Hscl}) and we rescale the position and momentum quadrature as $x=\sqrt{\frac{\omega_0}{\Omega}}x^{\prime}$ and $p=\sqrt{\frac{\omega_0}{\Omega}}p^{\prime}$. It leads to 
\begin{align}
\label{eq:Hsclpm}
H^\pm_{\textrm{scl}}(x,p)/\Omega= p^2/2+V^\pm_{\textrm{eff}}(x)/\Omega,
\end{align}
where the semiclassical effective potential~\cite{Ashhab:10,Ashhab:13,Bakemeier:12} reads
\begin{equation}
\label{eq:veff}
V^\pm_{\textrm{eff}}(x) /\Omega= \frac{1}{2}x^2 \pm
\frac{1}{2}\sqrt{1 + 2 g^2 x^2}.
\end{equation}

\begin{figure}
  \includegraphics[width=0.5\linewidth,angle=-90]{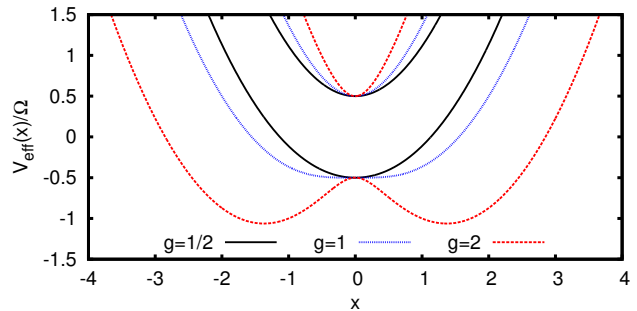}
\caption{(Color online) Representation of the effective potential
  $V^\pm_{\textrm{eff}}(x)/\Omega$ given in Eq.~(\ref{eq:veff}) for three
  different values of $g$; below the critical value
  ($g=1/2$), at the critical value ($g=1$)
  and above ($g=2$) with solid (black), dotted (blue)
  and dashed (red) lines, respectively. While the branch with positive
  sign always features a unique minimum at $x=0$, whose energy is
  $\Omega/2$, the negative or low-energy branch exhibits a double-well
  bifurcation (see main text for more details).}
\label{fig:pot}
\end{figure}
In the Fig.~\ref{fig:pot}, we present
the effective potential for three different characteristic coupling
strength values, namely, $g=1/2$, $g=1$
and $g=2$. First, the effective potential of the high-energy spin subspace $V^+_{\textrm{eff}}(x)$ always has a unique energy minimum at $x=0$ for any $g$, and its minimum energy is much larger than the extremal energies of $V^-_{\textrm{eff}}(x)$. As we are interested in the critical behavior in the spectrum of the low-energy spin subspace, in the following we do not concern ourselves with the high-energy spin subspace. On the other hand, the lower-energy effective potential $V^-_{\textrm{eff}}(x)$ has a unique minimum at $x=0$ for $g<1$. Then, for $g>1$, the energy minimum bifurcates to two local minima at $x=\pm
 \frac{1}{\sqrt{2}}\sqrt{g^2-g^{-2}}$, while $x=0$ becomes the local maximum. That is, the QPT of the Rabi model manifests itself in the semiclassical limit as a transition from a single-well to the double-well potential at the critical point $g=1$.
 
For $g>1$, the structure of a classical orbit in the phase space with a given energy $E$ abruptly changes when $E$ crosses $E_c=-\Omega/2$, i.e., $\varepsilon_c=-1$, which corresponds to the local maximum energy at the origin. For $E<E_c$, the classical orbits consists of two disconnected regions, localized in each of the double well. Since $H_{\textrm{scl}}(x,p,g)$ is invariant under $x\rightarrow -x$, the localized orbits in the double well potential indicate that the spontaneous symmetry breaking occurs. This is again the semiclassical manifestation of the doubly degenerate ground states as well as low-lying excited states with the broken parity symmetry~\cite{Hwang:15}. On the other hand, for $E>E_c$, the classical orbits consist only of a single region that is localized at the origin, just as in the case of $g<1$. This abrupt change in the phase space structure in the semiclassical limit is intimately related to the ESQPT~\cite{Stransky:14}, which we analyze in much more detail below. Note also that we have already witnessed the precursor of this abrupt change in the semiclassical phase space structure in the energy spectrum for $g>1$ [Fig.~\ref{fig:deg} (b)] as a sharp crossover; indeed, the phase boundary of the crossover coincides with the critical energy $E_c=-\Omega/2$ or, in terms of the dimensionless energy, $\varepsilon_c=-1$.

\subsection{Density of states}
\label{sec:iiib}
The semiclassical approximation of the quantum density of
states of a system with $f$ degrees of freedom is given by the
$f$-dimensional volume of the available phase space at a certain
energy $E$ and coupling strength $g$~\cite{Stransky:14},
which reads
\begin{align}
\nu(E,g) = \frac{1}{(2\pi)^f}\int d\vec{p}\, d\vec{q}\,
\delta\left[E-H_{\textrm{scl}}(\vec{q},\vec{p},g)\right].
\end{align}
The Rabi model has a single effective degree of freedom~\cite{Hwang:10}, $f=1$.  By using the semiclassical Hamiltonian $H^\pm_{\textrm{scl}}(x,p)$ in Eq.~(\ref{eq:Hsclpm}), the semiclassical density of states of the Rabi model reads
\begin{align}
\label{eq:scldos}
\nu(\varepsilon,g)&= \frac{1}{\omega_0\pi}\frac{\partial}{\partial \varepsilon} \int dx\,
dp \, \Theta\left[\varepsilon-p^2-x^2+\sqrt{1+2g^2x^2}\right] \nonumber
\\ &= \frac{\partial}{\partial \varepsilon} N(\varepsilon,g).
\end{align}
where $(x,p)$ are the rescaled coordinates, $\varepsilon= E/|E_c|=2E/\Omega$ is the rescaled energy and $N(\varepsilon,g)$ the accumulated number of states $N(\varepsilon,g)$. Note that $N(\varepsilon,g)$ is obtained as the total phase-space area
explored by the orbits of normalized energy $\varepsilon$ and dimensionless coupling strength
$g$. The accumulated number of states will be useful to address the critical behavior of certain observables, as we will see in Sec.~\ref{sec:iiic}. Making use of
Eq.~(\ref{eq:scldos}), the semiclassical density of states of the Rabi
model is given by
\begin{align}
\label{eq:scldos2}
\nu(\varepsilon,g) &= \frac{1}{\omega_0\pi} \int d x \, d p \, \Big( \frac{\delta
  \left[p-p_+ \right]}{\left|\partial_p H_{\textrm{scl}} (x,p,\lambda)2/\Omega
  \right|_{p=p_+} }+\nonumber \\ &\qquad\qquad\qquad\:\,+\frac{\delta \left[p-p_-
    \right]}{\left|\partial_p H_{\textrm{scl}} (x,p,\lambda)2/\Omega
  \right|_{p=p_-} } \Big) \nonumber \\ &= \frac{2}{\omega_0\pi}
\int_{x_{1}}^{x_2} \frac{dx}{\sqrt{\varepsilon-x^2+\sqrt{1+2g^2x^2}}},
\end{align}
where $p_{\pm}=\pm\sqrt{\varepsilon-x^2+\sqrt{1+2g^2x^2}}$ are the positive and negative roots of
$\varepsilon-2H_{\textrm{scl}}(x,p,\lambda)/\Omega=0$, and
the lower and upper limits of integration read
\begin{align}
x_{1}&=\sqrt{\varepsilon+g^2-\sqrt{g^4 +2\varepsilon g^2 +1}} \,
\Theta\left[\varepsilon_c-\varepsilon \right], \\
x_2 &=\sqrt{\varepsilon+g^2+\sqrt{g^4 +2\varepsilon g^2 +1}}.
\end{align}
It is clear from the previous expressions that for $g>1$
and $\varepsilon<\varepsilon_c=-1$ the classical orbits consist of two disconnected
regions, and that for either $g\leq 1$ or
$g>1$ and $\varepsilon\geq \varepsilon_c$, the classical orbits consist of a
connected region. 

\begin{figure}
\includegraphics[width=1.\linewidth,angle=-90]{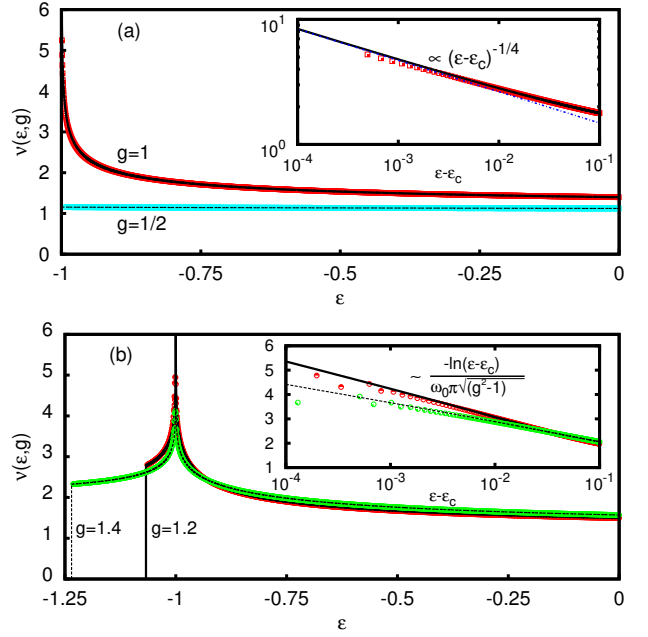}
\caption{(Color online) Semiclassical (lines) and quantum (points)
  density of states for different coupling strength at low energies,
  $\varepsilon\leq0$. Semiclassical results are obtained solving numerically the
  Eq.~(\ref{eq:scldos2}) and the quantum density of states corresponds
  to $\oo=10^3$, $\omega_0=1$ and $N=10$ (see Appendix~\ref{ap:b} for
  a detailed explanation to obtain it and its dependence on $N$). In
  (a) we consider $g=1/2$ (dashed line and light blue
  squares) and $g=1$ (solid line and red squares). In
  (b) we consider $g=1.2$ (solid line and red circles)
  and $g=1.4$ (dashed line and green circles). The
  vertical lines display the ground-state energy for the corresponding
  $\lambda$ value. In the insets we observe the diverging scaling
  behavior as $\varepsilon$ approaches to the critical energy $\varepsilon_c=-1$,
  either as a power law for $g=1$ (a), where the
  analytic solution is represented by a blue dot-dash line, or as a
  logarithmic singularity for $g>1$ (b) (see main text
  for details).}
\label{fig:dos}
\end{figure}
Now we derive an analytic expression for the density of states in two important limiting
cases: (i) $\varepsilon=\varepsilon_c$ at $g=1$ and (ii) $\varepsilon=\varepsilon_c$ for $g>1$. The former is concerned with the ground state QPT of the Rabi model as the $\varepsilon=\varepsilon_c$ is the ground state energy for $g=1$, while the latter concerns the ESQPT as the critical energy is much larger than the ground state energy. Let us start with the former, i.e., the ground state QPT, by denoting $\varepsilon=\varepsilon_c+\delta \varepsilon$ with $0<\delta \varepsilon\ll 1$. Since $\varepsilon>\varepsilon_c$, the lower
integration limit vanishes, $x_1=0$, while the upper limit can be
expanded in the lowest order in $\delta \varepsilon$ to give
\begin{align}
x_2&=\sqrt{\delta \varepsilon+\sqrt{2\delta \varepsilon}}=\left(2\delta \varepsilon \right)^{1/4}+\mathcal{O}\left(\delta\varepsilon^{3/4}\right).
\end{align}
The semiclassical density of states can be written as
\begin{align}
\nu(\varepsilon_c+\delta \varepsilon,g=1) &=\nonumber \\
 =\frac{2}{\omega_0\pi}
\int_{x_1}^{x_2} &\frac{dx}{\sqrt{\delta\varepsilon-1-x^2+\sqrt{1+2x^2}}}.
\end{align}
We carry out the integration in the leading order in $\delta \varepsilon$,
\begin{align}
\label{eq:qptdos}
&\nu(\varepsilon_c+\delta \varepsilon,g=1)\nonumber \\ &\approx \frac{2}{\omega_0\pi}\int_0^{(2\delta\varepsilon)^{1/4}}dx \left( \frac{1}{\sqrt{\delta\varepsilon -\frac{x^4}{2}}}+\mathcal{O}(x^6) \right) \nonumber \\
&= \frac{\Gamma(5/4)}{\Gamma(3/4)}\frac{2^{5/4}}{\omega_0\sqrt{\pi}}\delta \varepsilon^{-1/4},
\end{align}
where $\Gamma(m)$ is the Euler gamma function.  Hence, the
semiclassical density of states at the ground state QPT, $g=1$, diverges as 
\begin{equation}
\label{power}
\nu(\varepsilon,g=1)\propto
(\varepsilon-\varepsilon_c)^{-1/4}
\end{equation}
for $\varepsilon-\varepsilon_c\ll1$. This power-law divergence constitutes as a signature of the ground state QPT in the Rabi model in the semiclassical limit.

Now we consider the the ESQPT, i.e., for $g>1$ and for energy $\varepsilon$ that is close to $\varepsilon_c$. Again we denote the energy as $\varepsilon=\varepsilon_c+\delta \varepsilon$ with $0<\delta \varepsilon\ll 1$. We remind that for $g>1$, the critical energy  $\varepsilon_c$ is much larger than the ground state energy. The lower integration limit is $x_1=0$ as before, while the
upper one $x_2$ becomes
\begin{align}
x_2=\sqrt{2(g^2-1)}+\mathcal{O}\left(\delta \varepsilon\right).
\end{align}
Then, the density of states reads
\begin{align}
\label{eq:esqptdos_prev}
\nu(\varepsilon_c+\delta \varepsilon,g)&=\nonumber \\
=\frac{2}{\omega_0\pi}\int_{x_1}^{x_2}&\frac{dx}{\sqrt{-1+\delta \varepsilon-x^2+\sqrt{1+2g^2x^2}}}.
\end{align}
The previous expression has two possible singularities at $x_1=0$ and $x_2=\sqrt{2(g^2-1)}$  when $\delta
\varepsilon=0$. Therefore, we can
split the integral into two subintervals, namely,
$\int_{x_1}^{x_2}dx=\int_{x_1}^{x_m}dx+\int_{x_m}^{x_2}dx$.  However,
as we show in the Appendix~\ref{ap:a1}, the latter leads to a constant value $K$ when the integration is carried out. We choose $x_m$ to be small, $0<x_m\ll 1$, but greater than
$\delta \varepsilon$, so that we can resort to the Taylor expansion for
$x\ll 1$,
\begin{align}
\frac{1}{\sqrt{\delta \varepsilon-1-x^2+\sqrt{1+2g^2x^2}}}&\nonumber \\
= \frac{1}{\sqrt{\delta \varepsilon+(g^2-1)x^2}}+&\mathcal{O}(x^4).
\end{align}
This allows us to obtain the singular part of the density of states,
\begin{widetext}
\begin{align}
\nu(\varepsilon_c+\delta \varepsilon,g)&= \frac{2}{\omega_0 \pi}\int_0^{x_m}dx \left(\frac{1}{\sqrt{\delta \varepsilon+(g^2-1)x^2}}+\mathcal{O}(x^4)\right) +K \approx \frac{1}{\omega_0\pi\sqrt{g^2-1}}\ln\left(\frac{2x_m^2(g^2-1)}{\delta \varepsilon} \right)+K.
\end{align}
\end{widetext}
Hence, the semiclassical density of states diverges for $g>1$ at $\varepsilon=\varepsilon_c$ but differently compared to the ground state QPT
case. Namely, it shows a logarithmic singularity at $\varepsilon=\varepsilon_c$, rather than the power-law divergence shown in Eq.~(\ref{power}). Although the previous expression is only valid for $\varepsilon-\varepsilon_c >0$, the same behavior is found for $\varepsilon-\varepsilon_c< 0$ (see Appendix~\ref{ap:a2}). Therefore, for $g>1$, the semiclassical density of states shows the logarithmic divergence at $\varepsilon_c$ as
  \begin{align}
\label{eq:esqptdos}
\nu(\varepsilon,g>1)\sim
\frac{-\ln\left|\varepsilon-\varepsilon_c\right|}{\omega_0\pi\sqrt{g^2-1}}
\quad \textrm{for} \quad \left|\varepsilon-\varepsilon_c \right|\ll 1.
\end{align}
The logarithmic divergence at a critical energy for the broken symmetry phase demonstrates that the Rabi model exhibits an ESQPT.

We confirm our analytical expressions for the singular part of the density of states in the limiting cases by calculating the density of states numerically from the Eq.~(\ref{eq:scldos2}) for several representative values of $g$ [Fig.~\ref{fig:dos} (a) and (b)]. In both cases of the ground state QPT ($g=1$) and the ESQPT ($g>1$), the predicted power-law and the logarithmic divergence, respectively, shows excellent agreement with the numerically calculated density of states.

Finally, we corroborate the semiclassical analysis of the ESQPT with the numerical solution of the quantum Rabi model with a large, but finite, $\oo$ value. To
this end, we compute the quantum averaged density of states
$\bar{\nu}_q(\varepsilon,\lambda)$. Consider a window of energy spectrum consists of $N$ consecutive eigenstates, whose width is $\Delta \varepsilon$ and the energy in the middle is $\bar{\varepsilon}$. We calculate the quantum averaged density of states at an energy $\bar{\varepsilon}$ as $N/\Delta \varepsilon$. In the
Appendix~\ref{ap:b} we present a detailed explanation of the method to
compute $\bar{\nu}_q(\varepsilon,g)$ and discuss its dependence on the free parameter $N$. As an example, we choose $\Omega/\omega_0=10^3$ and $N=10$. The quantum averaged density of states agrees well with the semiclassical density of states [Fig.~\ref{fig:dos} (a) and (b)]. We note that the quantum density of states does not diverge but saturates at a certain value [Insets of Fig.~\ref{fig:dos} (a) and (b)]. This is due to the finite-frequency effect which smoothens out the singularity. Nevertheless, its scaling behavior close to $\varepsilon_c$ agrees with the semiclassical result.

We note that the Dicke model also exhibits the ESQPT with the logarithmic divergence in the first derivative of the density of states. Therefore, our finding shows that the ESQPT of the Rabi model manifests itself differently than in the Dicke model~\cite{Brandes:13,Bastarrachea:14}. This difference can be understood from the consideration of the number of effective degrees of freedom of the two models~\cite{Stransky:14}. While the low-energy physics of the Rabi model in the $\ool$ limit is effectively described by the single harmonic oscillator~\cite{Hwang:15}, the low-energy physic of the Dicke model in the thermodynamic limit is described by two harmonic oscillators where the additional oscillator represents the infinitely many TLSs~\cite{Emary:03}. As discussed in the Ref.~\cite{Stransky:14}, a local maximum in the phase space of the system with a single degree of freedom entails a logarithmic divergence in the density of states, while the saddle point in the phase space of a system with two degrees of freedom entails a logarithmic divergence of the first derivative of the density of states. We also note that the Lipkin-Meshkov-Glick model~\cite{Ribeiro:07,Ribeiro:08}, which has a single effective degree of freedom shows the logarithmic divergence in the density of states as in the case of the Rabi model.

\begin{figure}
\includegraphics[width=0.5\linewidth,angle=-90]{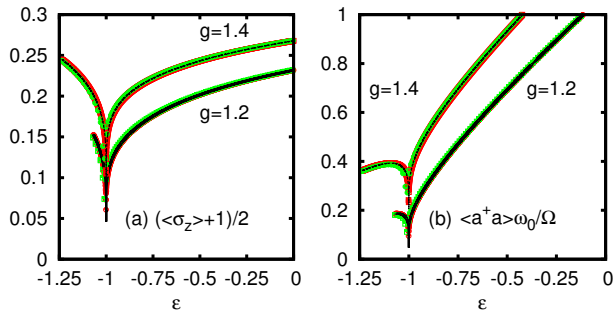}
\caption{(Color online) Comparison between quantum and semiclassical
  results, obtained from Eqs.~(\ref{eq:adagscl}) and~(\ref{eq:szscl}),
  for the rescaled photon number, $\braket{\adag
    a}\frac{\omega_0}{\Omega}$ in (a), and the TLS occupation,
  $\left(\braket{\sigma_z}+1\right)/2$ in (b) at
  $g=1.2$ and $g=1.4$ as a function of the normalized energy $\varepsilon$. The
  semiclassical result is depicted by a solid black line, while the
  quantum results are represented by red circles ($\oo=10^3$) and
  green squares ($\oo=10^2$), where each point corresponds to an
  expectation value of a particular eigenstate
  $\ket{\varphi_k^{\pm}}$. Note the singular behavior at $\varepsilon=\varepsilon_c=-1$.}
\label{fig:obs}
\end{figure}

\subsection{Signatures of excited-state quantum phase transitions in physical observables}
\label{sec:iiic}
Here we show that the singularity of the density of states discussed in the previous section leads to critical behaviors in observables~\cite{Caprio:08,Brandes:13,Perez-Fernandez:11}, opening up a possibility of an experimental observation of the ESQPT~\cite{Engelhardt:15}. The
semiclassical approximation to the expectation value of an
observable $\mathcal{A}$ can be obtained from~\cite{Brandes:13}
\begin{align}
\braket{\mathcal{A}}(\varepsilon,g)&=\frac{1}{\nu(\varepsilon,g)} \sum_{k,\pm} \bra{\varphi_k^{\pm}}\mathcal{A} \ket{\varphi_k^{\pm}}\delta(\varepsilon-\varepsilon_k^{\pm}) ,
\end{align}
where $\braket{\mathcal{A}}(\varepsilon,g)$ stands for the energy averaged
value of the observable $\mathcal{A}$ at energy $\varepsilon$ and coupling
strength $g$. If the Hamiltonian linearly depends on the observable $\mathcal{A}$ with a proportional constant $\beta$, i.e.,
$\mathcal{A}=\partial_{\beta} H$, the averaged value $\braket{\mathcal{A}}$ can be obtained using the Hellmann-Feynman theorem~\cite{Brandes:13}, that is,
\begin{align}
\braket{\mathcal{A}}(\varepsilon,g)&=-\frac{1}{\nu(\varepsilon,g)}\frac{\partial}{\partial \beta }N(\varepsilon,g),
\end{align} 
where $N(\varepsilon,g)=\int_{-\infty}^\varepsilon d\varepsilon' \, \nu(\varepsilon',g)$. Note that the dependence of $N(\varepsilon,g)$ on $\beta$ is not explicitly written. For the Rabi model, we have $a^\dagger a= \partial_{\omega_0} H$ and $\sigma_z= \partial_{\Omega/2} H$. Therefore, we obtain
\begin{align}
\label{eq:adagscl}
\braket{\adag a}(\varepsilon,g)&=-\frac{1}{\nu(\varepsilon,g)}\frac{\partial}{\partial \omega_0}N\left(\varepsilon,g\right), \\
\label{eq:szscl}
\braket{\sigma_z}(\varepsilon,g)&=-\frac{2}{\nu(\varepsilon,g)}\frac{\partial}{\partial \Omega}N\left(\varepsilon,g\right).
\end{align}

\begin{figure}
\includegraphics[width=0.57\linewidth,angle=-90]{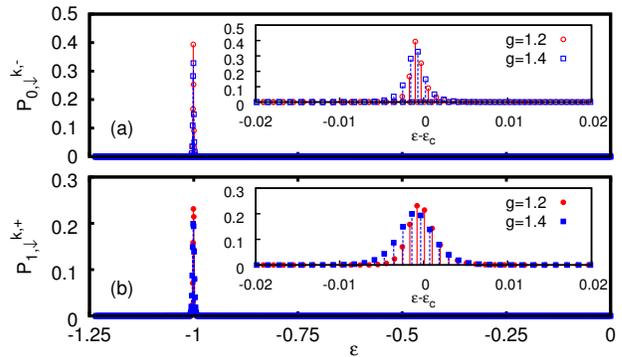}
\caption{(Color online) Probabilities $P_{0,\downarrow}^{k,-}$ (a) and
  $P_{1,\downarrow}^{k,+}$ (b) as a function of the normalized eigenstate-energy $\varepsilon$
  for a system with $\oo=10^3$ at two different coupling strengths,
  $g=1.2$ and $g=1.4$, depicted by
  circles (red) and squares (blue), respectively. Each point
  corresponds to a different eigenstate. The insets show a zoom close
  to the critical energy.}
\label{fig:vac}
\end{figure}

Both observables are directly related to the density of states and therefore the singularity in the density of state leads to the critical behavior in their mean-field value. We present the numerical results for the TLS population $\frac{1}{2}(\braket{\sigma_z}+1)$ [Fig.~\ref{fig:obs} (a)] and the rescaled photon number $\braket{\adag a}\frac{\omega_0}{\Omega}$ [Fig.~\ref{fig:obs} (b)], which show the critical behaviors at the critical energy $\varepsilon_c=-1$. Interestingly, we also observe precursors of
this critical behavior in the quantum expectation values, i.e., in
$\bra{\varphi_k^{\pm}}(\sigma_z+1)/2\ket{\varphi_k^{\pm}}$ and
$\bra{\varphi_k^{\pm}}\adag a\ket{\varphi_k^{\pm}}$, when
$\varepsilon_k^{\pm}=\varepsilon_c$ and $g>1$, provided by a large frequency
ratio $\oo$ [Fig.~\ref{fig:obs} (a) and (b)]. As an example, we have chosen $\oo=10^2$ and $10^3$. For the
larger values of $\oo$, we see the better agreement between the quantum and the
semiclassical results. We remark that this singular behavior is
present for any $g>1$ at the critical energy $\varepsilon_c$, and
therefore it is not constrained to a particular coupling strength
value.

The interesting feature of the ESQPT is that the eigenstates around
$\varepsilon_c$ for $g>1$ have a vanishing average population of the TLS and photon. In other words, for
eigenstates with negative parity, the probability of finding
$\ket{0,\downarrow}$, that is,
$P_{0,\downarrow}^{k,-}=\left|\left<0,\downarrow |\varphi_k^{-}
\right> \right|^2$ is maximal for some $k$ if $\varepsilon_k^{-}\approx
\varepsilon_c$. This can be understood as a localization of the wave
function around $x=0$~\cite{Caprio:08,Santos:15}. Indeed, in the
Fig.~\ref{fig:vac}(a), we represent $P_{0,\downarrow}^{k,-}$ for
$\oo=10^3$ at two different coupling strengths, $g=1.2$ and $g=1.4$ as
a function of the energy $\varepsilon_{k}^{\pm}$. The closer the energy
of the eigenstates to $\varepsilon_c$, the larger the
$P_{0,\downarrow}^{k,-}$ value. Clearly, for positive parity
eigenstates, this probability vanishes, $P_{0,\downarrow}^{k,+}=0$
since the state $\ket{0,\downarrow}$ belongs to the negative
parity. However, same conclusion can be drawn considering
$P_{1,\downarrow}^{k,+}$ for positive parity eigenstates, as one can
see in Fig.~\ref{fig:vac}(b).

\section{Conclusions}
\label{sec:iv}
In the present article, we have demonstrated that the second-order ground-state quantum phase transition (QPT) of the Rabi model is accompanied by an excited-state quantum phase transition (ESQPT) in the broken symmetry phase, in the sense that there exists a critical energy where the semiclassical density of states exhibits the logarithmic singularity and the semiclassical average values of observables show critical behaviors. The semiclassical analysis of the Rabi model has so far been mainly concerned with the critical behaviors at the ground state energy~\cite{Ashhab:13,Bakemeier:12}; but, here we have extended its scope to higher energy domain and both analytically and numerically demonstrated the presence of the criticality in the framework of the ESQPT. We also have shown that the precursors of the ESQPT appear in the fully quantum mechanical solution for a large but finite values of $\Omega/\omega_0$ with an excellent quantitative agreement, except the regularized singularity due to the finite-frequency effect.

The ESQPT has been understood to occur in a system with a few effective degrees of freedom arising from infinitely many system components~\cite{Stransky:14}. Our analysis however shows that the ESQPT also arises in a system with finite number of system components; moreover, the general classification of the ESQPT in terms of the number of effective degree of freedom is still valid in the finite system case. Our study that the Rabi model consisting only of a single oscillator and a two-level system exhibits the ESQPT adds another important aspect to the emerging field of quantum phase transition and critical phenomena in finite quantum systems~\cite{Hwang:15,Hwang:16}. Finally, we emphasize that the Rabi model, due to its ubiquity and
simplicity, may offer a promising model system to understand the ESQPT both theoretically and
experimentally.

\begin{acknowledgments}
This work is supported by an Alexander von Humboldt Professorship, the
EU Integrating Project DIADEMS, the EU STREP project EQUAM and the ERC Synergy grant BioQ. This work
was performed on the computational resource bwUniCluster funded by the
Ministry of Science, Research and Arts and the Universities of the
State of Baden-W\"urttemberg, Germany, within the framework program
bwHPC.
\end{acknowledgments}

\appendix
\section{Logarithmic singularity in the semiclassical density of states}
\label{ap:a1}
In the main text has been argued that the semiclassical density of
states, given in Eq.~(\ref{eq:scldos2}), undergoes a logarithmic
singularity for $g>1$ at $\varepsilon=\varepsilon_c$ plus some constant
value $K$. However, the Eq.~(\ref{eq:esqptdos}) is just a result of
the integration close to the origin $x=0$. Here we present a detailed
derivation of the second part of the integral, which will result in a
constant shift $K$.  As starting point we consider the semiclassical density
of states for a coupling constant $g>1$ and at an energy
$\varepsilon=\varepsilon_c+\delta \varepsilon$ with $0<\delta \varepsilon \ll 1$. Note
that the case for $\varepsilon=\varepsilon_c-\delta \varepsilon$ with $0<\delta
\varepsilon \ll 1$ is considered in the Appendix~\ref{ap:a2}. The density of
states, which is given in the Eq.~(\ref{eq:esqptdos_prev}), reads
\begin{align}
\label{eq:a1}
\nu(\varepsilon_c+\delta \varepsilon,g)&=\nonumber \\
=\frac{2}{\omega_0\pi}\int_{x_1}^{x_2}&\frac{dx}{\sqrt{\delta\varepsilon-1-x^2+\sqrt{1+2g^2x^2}}}, 
\end{align}
where $x_1=0$ and $x_2=\sqrt{2(g^2-1)}+\mathcal{O}(\delta \varepsilon)$. 
The previous integral can be split in two subintervals. The first subinterval, ranging from $x_1=0$ to $0<x_m\ll 1$, results in a logarithmic singularity as we have shown
in the main text (see Eq.~(\ref{eq:esqptdos})). Here we consider the
second subinterval, which again we split in two subintervals,
$\int_{x_m}^{x_2}dx=\int_{x_m}^{x_n}dx +\int_{x_n}^{x_2}dx$, being $x_m<x_n<x_2$ and $0<x_2-x_n\ll 1$. Note that the integral from $x_m$ to $x_n$ gives
just a constant since it does not involve any singular point, which we
denote $\tilde{K}$. To the contrary, the function to be integrated evaluated at
$x_2=\sqrt{2(g^2-1)}+\mathcal{O}(\delta \varepsilon)$ diverges as $\delta\varepsilon\rightarrow 0$.

Therefore, in order to analyze whether the Eq.~(\ref{eq:a1}) undergoes
a true singularity at $x_2$, we Taylor expand it around $x_2$. Thus,
we obtain
\begin{widetext}
\begin{align}
\label{eq:a2}
&\frac{2}{\omega_0\pi}\int_{x_m}^{x_2}\frac{dx}{\sqrt{\delta \varepsilon-1-x^2+\sqrt{1+2g^2x^2}}}=\tilde{K}+\frac{2}{\omega_0\pi}\int_{x_n}^{x_2}\frac{dx}{\sqrt{\delta \varepsilon-1-x^2+\sqrt{1+2g^2x^2}}}=\nonumber \\&=\tilde{K} +\frac{2}{\omega_0\pi}\int_{x_n}^{x_2}dx \left(\frac{1}{\sqrt{ \delta \varepsilon + \frac{2\sqrt{2}(g^2-1)^{3/2}}{2g^2-1}(x-x_2)}}+ \mathcal{O}((x-x_2)^2)\right)\approx \nonumber\\ &\approx \tilde{K}+\frac{2}{\omega_0\pi}\frac{(2g^2-1)}{\sqrt{2}(g^2-1)^{3/2}}\left(\sqrt{\delta\varepsilon+\frac{4(g^2-1)^2-2\sqrt{2}(g^2-1)^{3/2}}{2g^2-1}}-\sqrt{\delta \varepsilon}  \right).
\end{align}
\end{widetext}
For $\delta \varepsilon=0$, i.e., $\varepsilon=\varepsilon_c$ and $x_n=\sqrt{2(g^2-1)}-\delta x$, the Eq.~(\ref{eq:a2}) reads
\begin{align}
\frac{2}{\omega_0\pi}\int_{x_m}^{x_2}&\frac{dx}{\sqrt{\delta\varepsilon-1-x^2+\sqrt{1+2g^2x^2}}}\approx \nonumber \\ &\approx\tilde{K}+\frac{2}{\omega_0\pi}\frac{2^{1/4}\sqrt{\delta x(2g^2-1)}}{(g^2-1)^{3/4}}=K,
\end{align}
which is clearly analytic for any $g>1$ when $\delta\epsilon\rightarrow 0$. Hence, the semiclassical
density of states for $g>1$ does not feature a real singularity at
$x_2$. In short, we have shown that second subinterval of the integral
in Eq.~(\ref{eq:a2}) results in a constant value $K$, which just
produces a shift on $\nu(\varepsilon_c+\delta \varepsilon,g>1)$ and
consequently it does not affect the logarithmic divergence as $\delta\varepsilon\rightarrow 0$ for $0<\delta\varepsilon
\ll1$, given in Eq.~(\ref{eq:esqptdos}).

\subsection{Semiclassical density of states for $g>1$ and $\varepsilon=\varepsilon_c-\delta\varepsilon$}
\label{ap:a2}
Here we show the logarithmic singularity in the semiclassical density of states for $g>1$ and $\varepsilon=\varepsilon_c-\delta \varepsilon$, being $0<\delta \varepsilon\ll 1$. In this case, the integration limits $x_1$ and $x_2$ can be approximated as
\begin{align}
x_1&=\sqrt{\frac{\delta \varepsilon}{g^2-1}}+\mathcal{O}(\delta \varepsilon) \\
x_2&=\sqrt{2(g^2-1)}+\mathcal{O}(\delta \varepsilon).
\end{align}
Then, the density of states results in
\begin{align}
  \nu(\varepsilon,g)&=\frac{2}{\omega_0\pi}\int_{x_1}^{x_2}dx\left(\frac{1}{\sqrt{(g^2-1)x^2-\delta \varepsilon}}+\mathcal{O}(x^4)\right) \\
&\approx \frac{1}{\omega_0\pi\sqrt{g^2-1}}\left(\ln(8(g^2-1)^2)-\ln(\delta \varepsilon)\right).
\end{align}
Therefore, $\nu(\varepsilon,g>1)$ also diverges logarithmically when $\varepsilon<\varepsilon_c$ and consequently, 
\begin{align}
\nu(\varepsilon,g)\sim -\frac{\ln\left|\varepsilon-\varepsilon_c\right|}{\omega_0\pi\sqrt{g^2-1}} \quad \textrm{for} \quad \left|\varepsilon-\varepsilon_c \right|\ll 1.
\end{align}
\begin{figure}
  \includegraphics[width=0.5\linewidth,angle=-90]{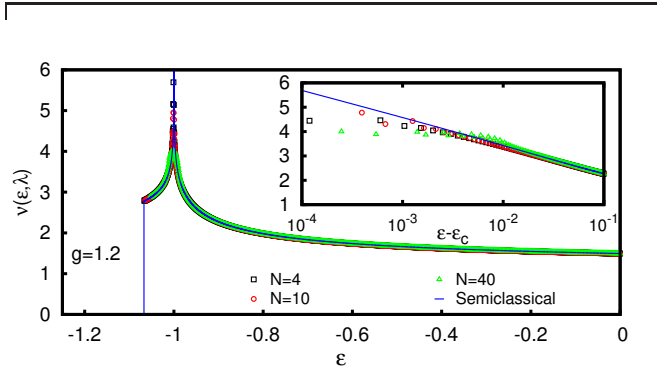}
\caption{(Color online) Comparison between the semiclassical (solid
  line) and quantum averaged density of states (points) for different
  values of $N$, for $\oo=10^3$ and $g=1.2$. The
  squares (black), circles (red) and triangles (green) correspond to
  $N=4$, $10$ and $40$, respectively. The inset shows the scaling
  behavior close to the critical energy.}
\label{fig:Ncomp}
\end{figure}

\section{Quantum averaged density of states}
\label{ap:b}

Since we are interested in a comparison between the semiclassical and
the quantum density of states, it is mandatory to perform an average
of the latter, which for a set of eigenstates with energies
$\varepsilon_k^{\pm}$ reads $\nu_q(\varepsilon,g)=\sum_{k,\pm}
\delta\left(\varepsilon-\varepsilon_k^{\pm} \right)$.  Therefore, we
need to compute a quantum averaged density of states denoted by
$\bar{\nu}_q(\varepsilon,g)$, which is obtained as follows. For the
$i$-th eigenstate, we obtain the energy difference $\Delta
\varepsilon(i,N)=\varepsilon_{i+N}-\varepsilon_{i}$ and the middle
energy
$\bar{\varepsilon}(i,N)=(\varepsilon_{i+N}+\varepsilon_{i})/2$. Hence,
the quantum averaged density of states is given as $N/\Delta
\varepsilon(i,N)$ at the energy $\bar{\varepsilon}(i,N)$, that is,
$\bar{\nu}_q(\bar{\varepsilon}(i,N),\lambda)=N/\Delta
\varepsilon(i,N)$. In this way, the only free parameter is $N$, the
width of the window where the average is performed.  Qualitatively,
the size of the window has to be large enough to provide a reliable
average but still small to prevent excessively blurred
outcomes. Quantitatively, this can be done comparing different values
of $N$ for the same set of eigenenergies $\varepsilon_k^{\pm}$. In the
Fig.~\ref{fig:Ncomp}, we represent the quantum averaged density of
states for $\oo=10^3$ at $g=1.2$ for different values of $N$, namely,
$N=4$, $10$ and $40$. For a large window size ($N\gtrsim 40$) the
behavior is smoother and a numerical artifact appears close to the
critical energy, which is visible in the inset of
Fig.~\ref{fig:Ncomp}. On the other hand, for smaller values of $N$
there are no significant differences between them, meaning that they
represent a good average. Finally, we emphasize that, even though the
specific value of $\bar{\nu}_q(\varepsilon,g)$ depends on $N$, the
semiclassical density of states is faithfully reproduced, as well as
the scaling to close to the critical energy, provided by a reasonable
$N$. Hence, in order to verify the agreement between the semiclassical
and quantum results, we choose an intermediate value of $N$, i.e.,
$N=10$ for the results presented in the main text, but the conclusions
do not change for any other $N$.

\bibliographystyle{apsrev4-1.bst}

\bibliography{esqpt_rabi}

\end{document}